# Probabilistic Method for Optimizing Submarine Search and Rescue Strategy Under Environmental Uncertainty

Runhao Liu[1][✉], Ziming Chen[2] and Peng Zhang[3]

[1] Polytechnic Institute, Zhejiang University, Hangzhou 310015, China
runhaoliu@zju.edu.cn
[2] School of Social and Public Administration, East China University of Science and Technology, Shanghai 200237, China
[3] School of Mathematical Sciences, Zhejiang University, Hangzhou 310058, China

**Abstract.** When coping with the urgent challenge of locating and rescuing a deep-sea submersible in the event of communication or power failure, environmental uncertainty in the ocean can't be ignored. However, classic physical models are limited to deterministic scenarios. Therefore, we present a hybrid algorithm framework combined with dynamic analysis for target submarine, Monte Carlo and Bayesian method for conducting a probabilistic prediction to improve the search efficiency. Herein, the Monte Carlo is performed to overcome the environmental variability to improve the accuracy in location prediction. According to the trajectory prediction, we integrated the Bayesian based grid research and probabilistic updating. For more complex situations, we introduced the Bayesian filtering. Aiming to maximize the rate of successful rescue and costs, the economic optimization is performed utilizing the cost-benefit analysis based on entropy weight method and the CER is applied for evaluation.

**Keywords:** Submarine Detection, Monte Carlo, Grid Analysis, Bayesian Method, Economic Optimization.

## 1 Introduction

The ocean is a treasure trove of myriad potential resources, as 70% of the Earth's surface is encompassed by the ocean [1]. With the development of submersibles, it is more and more possible for humans to arrive at complex deep sea environments to develop the underwater resources [2]. In this situation, deep-sea exploration has rapidly evolved into a prominent sector of marine resource development, attracting significant public interest. Despite these technological advancements, significant safety concerns persist, as evidenced by recent high-profile incidents involving submersibles losing contact with their host vessels [3]. Consequently, the demand for robust localization and rescue strategies has intensified to ensure the safety of deep-sea resource development.

Recently, several methods for ocean resource development are proposed. Saleem designed an AI-driven routing protocol for underwater communication in marine energy exploration [4]. Zhou et al. proposed a deep learning-based object detection method for resources development in the complex underwater environment [5]. These researches contributed profitable ideas for exploring underwater resources using submarines. However, an efficient search and rescue system is needed to overcome the accident of submarine [6].

The inherent complexity and unpredictability of the underwater environment pose substantial challenges to precise localization and timely rescue operations, which could be influenced by numerous environmental factors [7]. Several deterministic methods are developed for the prediction of submarines' location. However, early methods like empirical models with parameterized formulas [8, 9], dynamic integration [10], linear drift models [11, 12] assume fixed and simplistic environmental conditions, which might lead to the failure of providing accurate estimations under stochastic influences.

Meanwhile, probabilistic methods attempt to address these uncertainties, but they still have notable limitations. Kalman filters can be significantly restrictive in the non-linear and non-Gaussian marine environments [13]. Particle filters required extensive computational resources while this method has limited reliability and robustness in complex scenarios [14]. Yao et al. applied expectation-maximization (EM) to search underwater targets [15]. Li et al. utilized the RRT algorithm and neural network for the compatibility of multi-objective search [16]. However, these methods are all low in efficiency when search targets dynamically move with uncertainty.

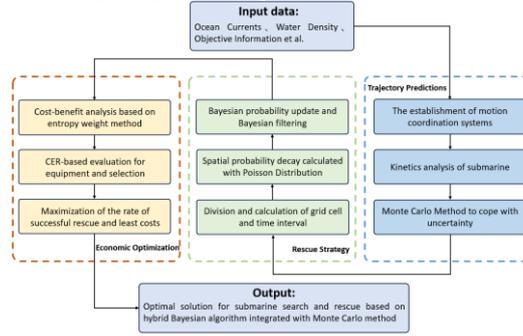

**Fig. 1.** The flowchart of proposed model.

To cope with limitations of aforementioned methods in searching and rescuing the submarine, this paper introduces a Bayesian Monte Carlo framework specifically designed to address these critical limitations, as shown in Fig. 1. Above all, we analyzed the motion behavior of the target submarine under different conditions. To solve the uncertainty caused by complex environment, the Monte Carlo is introduced. Then, a grid-based probabilistic model based on Poisson distribution integrated with Bayesian method is conducted to optimize search paths dynamically. Besides, economic optimization is also conducted to save the cost. Our main contributions are listed below:

(1) We ensembled the kinetics analysis and the Monte Carlo method. Instead of single physical models, the proposed method simulated the influence of currents with the Monte Carlo method to provide a more precision trajectory predication.

(2) On the basis of the trajectory prediction, a Poisson distribution-based grid probability search is designed to realize the information-driven optimization of the dynamic search strategy. Herein, the initial search probability is constructed based on Poisson distribution and the continuous correction of regional priority is conducted by Bayesian while a Bayesian filtering is also used for multiple objective tasks. Our method enhanced search efficiency and adaptive capabilities.

(3) Under the premise of ensuring the effectiveness of the search and rescue strategy, we further introduced the cost-benefit analysis model (CBA) to optimize the strategy through entropy weighting method and CER evaluation.

## 2 Problem Description and Method

As shown in Fig. 2, this paper aims to accurately predict the trajectory and final location of a lost submersible under conditions of uncertainty caused by complex marine environments. Thus, the correlation analysis of the environmental elements, including ocean currents, water density, and geographic locations, should be conducted. Then, the kinetic characters of the objective lost submarine are needed to be analyzed for trajectory prediction in scenarios with uncertainty. We assume the lost submarine is in the worst situation for the simultaneous loss of engine power and communication. For this purpose, we adopt a probabilistic modeling framework that integrates Monte Carlo simulation and Bayesian inference. After that, it is an inevitable requirement to propose an optimal search and rescue strategy for the shortest time required for rescue. Additionally, cost control is necessary while ensuring a high success rate in search and rescue missions. Therefore, economic analysis is performed for the comparison of rescue mission and cost based on the entropy weight method and CER analysis.

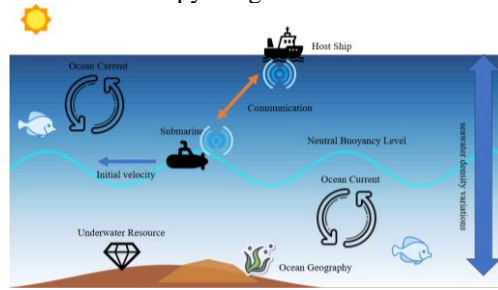

**Fig. 2.** Demonstration of problem background.

## 3 Mathematical Modeling

### 3.1 Search and Rescue Strategy Utilizing Poisson Distribution based Grid Probability Search

In emergency situations, the submersible's motion state can typically be categorized into two scenarios. First, when the neutral buoyancy point is located above the seabed, the submersible drifts slowly under the influence of ocean currents as gravity and buoyancy reach equilibrium. Second, when the neutral buoyancy point lies below the seabed, the submersible gradually sinks. Herein, deterministic values for ocean current characteristic are adopted. Then, by establishing the motion reference system, the underwater motion in the unpowered situation is considered. The fixed uniformly distributed mass assumption is adopted. The influence of currents is temporarily ignored because the size of the submersible is much smaller than the characteristic length of the currents. Eventually, the equations of motion of the target are obtained. Undynamic motion of submarine analysis based on the Monte Carlo can be conducted.

The parallel grid search is one of the most common technologies used for modern search and rescue missions. Based on parallel grid search, as last data before the submarine lost input, the grid search can be applied into the search area. In grid search, the search area is divided into smaller cells assigned with a probability percentage of the likelihood that the target submarine is located with the cell. Therefore, we note $G_s$ as the size of grid. With the maximum length of the horizontal axis denoted as $x_{max}$. Then, we define the $M_g = \frac{x_{max}}{G_s}$. Thus, the grid label of any point is expressed as $N_g$.

$$N_g = INT(\frac{x}{G_s}) + M_g \times INT(\frac{y}{G_s})$$ (1)

Based on the (1), the coordination of the center position $(x_G, y_G)$ is obtained in.

$$\begin{cases} x_G = (N_g \% M_g) \cdot G_s + \frac{G_s}{2} \\ y_G = INT(\frac{N_g}{M_g}) + \frac{G_s}{2} \end{cases}$$ (2)

We assume that the search time is within 30min, based on the parallel grid search, the grid size is calculated as (3).

$$G_s = \frac{v \times t}{Turns} = (W_s) \times Turns - (W_s \times O_l) \times (Turns - 1)$$ (3)

Wherein, the $W_s$ is the Swath Width and the $O_l$ represents the overlap. Based on this grid with the size of about $300m \times 300m$, the regional rasterization is conducted.

With the guidance of grid search, we discretize the rescue region into grids and assign each grid a probability of containing the target. Initially, the submersible is assumed to be most likely located in the predicted central grid, and the probability of locating it decreases with increasing distance from that grid. As the distance is set as 4, the probability of finding the missing underwater target is (4).

$$P_f = P(\chi = 4) \cdot \frac{1}{N_4}$$ (4)

This spatial probability decay is modeled using a Poisson distribution, where the intensity parameter $\lambda$ governs how sharply the probability falls off. Based on the Manhattan distance, the $k$ in the Poisson distribution function is obtained. However, the uncertainty of the submersible's actual location increases due to the randomness of ocean environment. Thus, it can be assumed that the ratio of the probability is integrated with time. Then, the (5) is obtained. Wherein, the minimum necessary time for the rescue vessel to get ready. The $m_p$ is a constant value that have a direct impact on the initial successful rate of rescue. We introduced a constant value $t_0$ to control that the vessels' preparation time is the same.

$$\lambda = m_p + \ln \frac{t}{t_0} \tag{5}$$

To improve search efficiency over time, we apply Bayesian theory to update the probability distribution after each search interval. Since the grid size is calculated utilizing the full calculation competence of the ship, the time could also be divided into multiple time intervals. Note the $t_i$ as the necessary time to complete single region search, the rule for time division is shown in (6).

$$[t_0 + n \cdot t_i, t_0 + (n+1) \cdot t_i] \tag{6}$$

Once a region has been searched without success, we assume the probability of finding the submersible in that region drops to 0. In this method, the prior probability is updated to a posterior probability that reflects the remaining unexplored areas. This dynamic updating process ensures that rescue efforts focus on the most promising locations as new information is gathered during the search.

In scenarios involving multiple submersibles or more complex motion patterns, we further introduce a Bayesian filtering model to estimate the system state. This method updates the target's position by combining prior estimates with new observations at each step. Integrated with our kinetics model, it contributes to track the submersible more accurately over time and allows dynamic adjustment of the search path and sonar range, making the search more focused and efficient.

### 3.2 Economic Model based on Cost-Benefit Analysis and Entropy Weight Method

In real-world search and rescue operations, equipment selection can't rely solely on performance indicators. Practical constraints such as budget limitations, equipment availability, maintenance complexity, and environmental adaptability must also be considered. To systematically balance these trade-offs, we adopt a CBA model.

We define a matrix $X_0$ including various performance indicators and cost indicators of the involving equipment. Then, according to the standardization formula $z_{ij}$, as shown in (7), the standardization matrix $Z_0$ is acquired, which is applied to calculate the probability matrix $P$. In the matrix $P$, each element $p_{ij}$ is calculated as (8). The $i$ th entropy value $e_j$ is (9).

$$z_{ij} = \frac{x_{ij}}{\sqrt{\sum_{i=1}^{n} x_{ij}^2}} \tag{7}$$

$$p_{ij} = \frac{z_{ij}}{\sum_{i=1}^{n} z_{ij}} \tag{8}$$

$$e_j = -\frac{1}{\ln n} \sum_{i=1}^{n} p_{ij} \ln(p_{ij}) (j=1,2,...,4) \tag{9}$$

The range of the $e_j$ is [0,1]. With the $e_j$ increased, the differentiation degree of index $j$ is also greater. Thus, we can derive more information. The value of corresponding weight $W_j$, defined as (10) should also be increased.

$$W_j = \frac{1-e_j}{\sum_{j=1}^m (1-e_j)}, j = 1,2,...,4 \tag{10}$$

For representing the benefits $E_i$ and costs $C_i$ of each rescue device, the (11) is conducted. Wherein, the $C_i$ consists of the cost of purchase (C1) and maintenance (C2) of the equipment. Then, the evaluation score $CER$ for each device can be represented by E and C, as shown in (12).

$$\begin{cases} E_i = \sum_{j=1}^m W_j P_{ij} \\ C_i = \frac{C_{total(i)}}{\sum C} \end{cases} \tag{11}$$

$$CER = \frac{E}{C} \tag{12}$$

For the stand condition, D, S and F, the benefits $E$ is shown as (13).

$$E = W_1 D + W_2 S + W_3 F \tag{13}$$

In a rescue mission, both of the equipment for detection and rescue is needed. To measure the performance, we introduced the index of stability, which might be involved with the designed and materials. Feasibility is also the index of stability. These parameters are ranged from 0 to 1.

## 4 Results and Discussion

### 4.1 Data Preparation and Relevant Parameters Analysis

For this research, we selected the Ionian Sea as the study case. As for current data, the Copernicus Marine Data provided by the Copernicus Programme is indexed. The Copernicus Programme is the European Union's earth observation program, which is an open, public, and nonprofit program [17]. The data of water density in specific regions is found in the National Centers for Environmental Information (NCEI), a U.S. government agency that manages one of the world's largest archives of atmospheric, coastal, geophysical, and oceanic data [18]. Geophysical data can be retrieved by the General Bathymetric Chart of the Oceans (GEBCO). The GEBCO is an organization supported by the United Nations, whose aim is to provide the most authoritative publicly available bathymetry of the oceans worldwide [19].

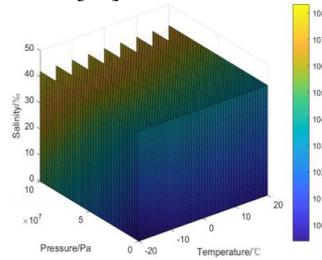

**Fig. 3.** Seawater salinity plotted against pressure and temperature

In order to construct a more accurate submersible motion prediction and localization model, we systematically analyzed key hydrographic information of the Ionian Sea. Fig. 3 depicts the coupling relationship between seawater salinity and temperature and pressure, showing the changing pattern of salinity at different depths. It can be seen that the density of seawater varies nonlinearly with depth, and the change process is driven by salinity, temperature and pressure. In addition, at the same depth, the density distribution shows relative consistency in the latitudinal direction, while the density of surface water increases with increasing latitude.

### 4.2 Kinetics Analysis of Submarine

The real marine environment exhibits a high degree of randomness, which is far from the explanation of neutral buoyancy. Thus, the Monte Carlo method of stochastic simulation is performed for the deep-sea target position prediction to overcome the substantial decline in the accuracy of trajectory predictions.

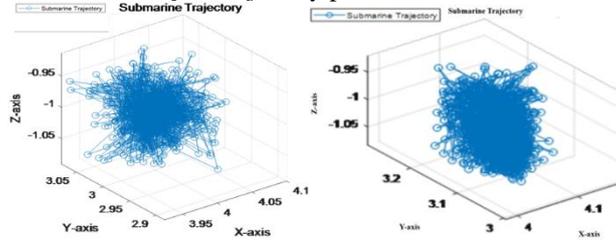

**Fig. 4.** The predicted (left) and initial velocity (right) trajectory map.

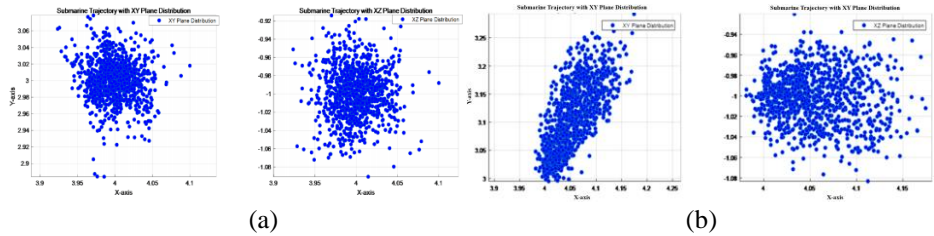

**Fig. 5.** The predicted (a) and initial velocity (b) trajectory point distribution with the projection of XOY and XOZ.

We introduced a total of 1,000 random particles to simulate the stochastic disturbances of ocean currents. As the initial state of the submersible is defined, the current disturbance factor is introduced with a velocity of [0.05, 0.30]. As for grid search, we set the $t_0$ as 20 min and the $t_i$ is set as 30 min. Then, the value of $m_p$ and $n$ is 0.2 and 10, respectively. Fig. 4 to Fig. 7 show the motion trajectories and position distributions of the submersible under different initial conditions. In Fig. 4, the trajectory and motion map of the submersible demonstrates its motion path under the influence of ocean current speed perturbation, especially under the initial position of [4, 3, -1] and zero initial velocity. When the initial velocity is 0, the motion trend of the submersible is mainly affected by the current perturbation, resulting in a maximum offset of its position in the horizontal plane of about 0.3 km. Fig. 5 (a) shows the distribution of the diver's trajectory points in the XOY and XOZ planes, which further revealed the mechanism by which the ocean currents affect the horizontal and vertical displacements of the diver. The horizontal initial velocity is considered in Fig. 5 (b). In this situation, the submersible is shifted in the direction of the initial velocity, with a smaller deviation in the horizontal displacement.

The probability distribution of the submersible position based on the results of Monte Carlo simulation is demonstrated in Fig. 6, which predicts the potential position of the submersible at different time points. Based on repeated simulations, it is concluded that the largest probability of the drop point occurred at the geometric distance of nearly 1 km from the failure start position. With an initial velocity [0.1, 0.2, 0], the submarine is more likely to move in the direction of the initial velocity. With the increase of time, the offset position in the horizontal direction is increasing.

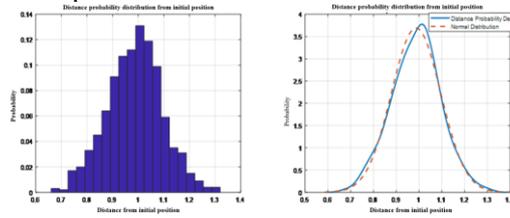

**Fig. 6.** Probabilities and probability densities of fallout distributions.

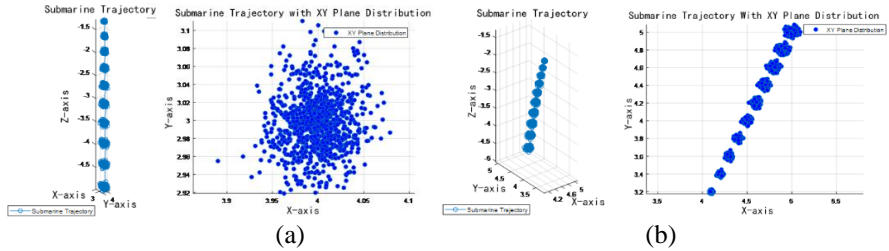

**Fig. 7.** Sinking trajectory and horizontal projection without (a) and with (b) horizontal initial velocity.

With the initial position set as [4, 3, -1], the acquired velocity matrix is [0, 0, 3] as the predicted location is in Fig. 7 (a). Fig. 7 (b) shows the results in the scenarios with initial velocity matrix [1, 2, -3] and the value of acceleration is set as 2. Caused by little frictional resistance of seawater, the dynamic change of the horizontal direction velocity is small during the sinking time while the horizontal velocity is almost merely influenced by current perturbation.

### 4.3    Search and Rescue Plan

As illustrated in Fig. 8, the strategy for search and rescue is formulated based on the submersible's buoyancy status and initial velocity. If the submersible is neutrally buoyant and lacks horizontal velocity, the search begins with a 1 km radius around the initial position, expanding over time. If it possesses initial horizontal velocity, sonar devices are deployed along the projected motion direction based on the estimated delay between failure and response. The maximum operating depth of is known about 4 km. In sinking scenarios, where the submersible descends due to gravity, deep-sea submersibles are used to search within the predicted motion direction (within 0.3 to 0.5 km) and sinking time. Multiple sonar units are coordinated to expand coverage and improve detection precision.

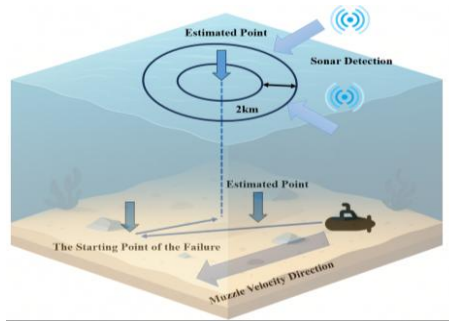

**Fig. 8.** Illustration of search schematic.

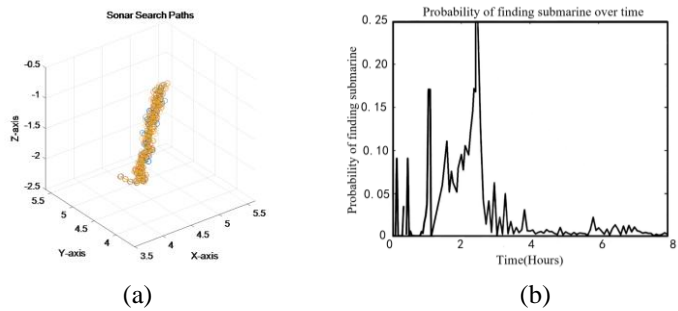

(a)                                      (b)

**Fig. 9.** The predicted search trajectory of the sonar (a) and relationships between allowed time and probability of successful searching (b).

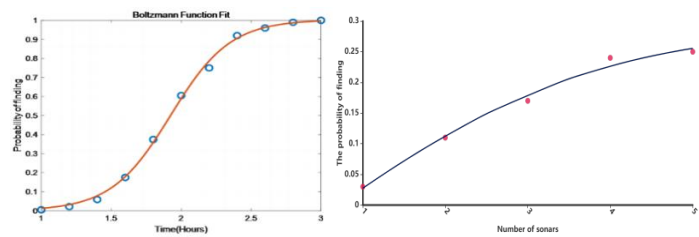

**Fig. 10.** The result of Boltzmann function fit of the relationship between: The allowed time and probability of successful searching for submersibles (left). The number of sonars and probability of successful searching for submersibles (right).

We conducted repeated simulations and plotted the search trajectory of the sonar, as shown in Fig. 9 (a). Then, based on the initial position at the time of the accident and its search range of 0.2 km to the left and right, i.e., [4.8, 5, -1] and [4.2, 5, -1], we configured the sonar devices. It is founded in Fig. 9 (b) that the probability of finding the submersible is greatest about 2 hours after the failure, while the probability of successfully finding the submersible decreases significantly 8 hours after the accident.

We conducted simulations to identify the optimal equipment combination. Starting with an initial position of [4, 5, -1] and velocity [-0.1, -0.2, -0.3], we set the sinking acceleration to 0.2 and used three sonar devices. The sonar search trajectory was plotted, and the relationship between time and success rate was calculated. After varying the number of sonars, we derived a relationship curve between the number of sonars and the probability of finding the submersible, as shown in Table 1. Based on this curve, the cumulative probability is calculated based on the functional relationship between probability and time by using the inhomogeneous point selection method and fitted to the Boltzmann function curve, and the fitting results are shown in Fig. 10, which confirms the correlation between time and search efficiency.

**Table 1.** The relationship between the number of sonobuoys and successful probability.

| Number of sonobuoys | 1 | 2 | 3 | 4 | 5 |
|---|---|---|---|---|---|
| Probability of finding the subma-rine | 2.87% | 11.23% | 17.34% | 24.00% | 24.69% |

### 4.4   Economic Analysis

**Table 2.** Instruments and parameters of relevance.

| Instruments | Function | Purchase Cost (C1) | Maintenance Cost (C2) | Detection range (D) | Stability (S) | Feasibility (F) | CER |
|---|---|---|---|---|---|---|---|
| Tritech SEK SK150 Side Scan Sonar | Detection | $7000 | $1100 | 200 m | 0.8 | 0.8 | 1.438 |
| FIFISH PRO V6 PLUS Underwater Exploration Drone | Detection | $6000 | $1600 | 150 m | 0.8 | 0.8 | 1.429 |
| REMUS Underwater Machines | Detection | $15000 | $3000 | 150 m | 0.9 | 1.0 | 0.743 |
| BlueComm 200 underwater communications equipment | Detection | $11500 | $1580 | 100 m | 0.7 | 0.8 | 0.804 |
| deep-diving survival boat | Rescue | $500000 | $19200 | 1000 m | 0.8 | 0.8 | 2.367 |
| Underwater rescue by divers | Rescue | $6000 | $1600 | 150 m | 0.7 | 0.6 | 1.555 |
| Zodiac Milpro SRMN 600 lifeboat | Rescue | $15000 | $3000 | 150 m | 0.7 | 0.9 | 0.873 |

We conducted the market survey and calculated the cost data associated with several common types of relevant equipments and Cost-Effectiveness Ratio (CERS) are calculated according to it, the result is shown in Table 2. Based on the CER values, the deep-diving survival boat, with a CER of 2.367, is the optimal choice for rescue, while the SK150 sonar, with a CER of 1.438, offers the best performance for detection. The optimal strategy suggests using four sonar devices for the best balance between coverage and efficiency. This strategy maximizes search effectiveness while minimizing costs by considering the characteristics and cost-effectiveness of each device. Based on the cost-benefit analysis of different equipment, we select underwater exploration drone as the primary surface rescue equipment owing to their lower cost and higher efficiency, especially for surface areas. For underwater search and rescue, we chose side-scan sonar (SSS) as the primary equipment.

## 5  Conclusion

In this work, a hybrid framework integrated with physical analysis and Monte Carlo method is established to predict the trajectory of lost submarine. Based on the calculation of motion and environmental uncertainties, the Monte Carlo-based prediction model is introduced to estimate the spatial probability distribution of the submersible's position. With the simulation of Monte Carlo, high uncertainty of ocean environment is solved with repeat simulation. Thus, the uncertain problem is converted into certainty. Then, a grid-based search strategy is developed using Poisson distribution to guide initial search. In this process, the Bayesian theory is applied to iteratively update the search probability as new information becomes available. To enhance search accuracy in complex scenarios involving multiple targets, we further proposed a Bayesian filtering method. This strategy realizes continuously update of initial search priorities through Bayesian inference as new search data is collected. This dynamic adjustment allows the search strategy to focus on the most promising areas, significantly increasing the likelihood of successful rescue within limited time and resources. The realize the highest rate of successful search and rescue with the optimal cost, we conducted a cost-benefit analysis model to evaluate and select the most efficient equipment. With the CER of 2.367, the deep-diving survival boat is the best choice for rescue while the SK150 prevails with the CER of 1.438. As for the placement of sonar, a number of 4 is the most comparison. Finally, an optimal rescue strategy with characteristic of various devices fully considered is formed to realize the maximum search effect with least costs.

**Acknowledgments.** This research was supported by National Key Research and Development Program of China (2024YFC2511003).